\documentclass[%
 reprint,
superscriptaddress,
 amsmath,amssymb,
 aps,
pra,
]{revtex4-2}

\usepackage{graphicx}
\usepackage{dcolumn}
\usepackage{bm}

\begin{document}

\preprint{APS/123-QED}

\title{STEM EBIC as a Quantitative Probe of Semiconductor Devices}

\author{Sebastian Schneider}
\email{sebastian.schneider2@tu-dresden.de}
\affiliation{Dresden Center for Nanoanalysis, cfaed, TUD University of Technology Dresden, 01069 Dresden, Germany}

\author{Sebastian Beckert}
\affiliation{Dresden Center for Nanoanalysis, cfaed, TUD University of Technology Dresden, 01069 Dresden, Germany}

\author{René Hammer}
\affiliation{Point Electronic GmbH, 06120 Halle (Saale), Germany}

\author{Markus König}
\affiliation{Max Planck Institute for Chemical Physics of Solids, 01187 Dresden, Germany}

\author{Grigore Moldovan}
\affiliation{Point Electronic GmbH, 06120 Halle (Saale), Germany}

\author{Darius Pohl}
\affiliation{Dresden Center for Nanoanalysis, cfaed, TUD University of Technology Dresden, 01069 Dresden, Germany}

\date{\today}

\begin{abstract}
Electron beam–induced current (EBIC) imaging in the scanning transmission electron microscope (STEM), STEM-EBIC,  provides direct access to carrier transport at the nanoscale. While well established in bulk SEM geometries, its application to thin TEM lamellae remains largely unexplored. Here, we present a systematic STEM-EBIC study of silicon photodiode lamellae prepared by gallium and xenon focused ion beam (FIB) milling. We directly visualize the p–n junctions in thin cross sections and extract effective diffusion lengths for electrons and holes as a function of local thickness. The values are orders of magnitude smaller than those obtained by SEM-EBIC on bulk silicon, reflecting pronounced surface recombination and FIB-induced surface modifications. Current–voltage measurements further reveal severe deviations from the expected diode-like behavior, which we attribute to ohmic metal–semiconductor contacts in the emasurement setup. Our analysis establishes STEM-EBIC as a quantitative probe of carrier transport in nanoscale devices. \end{abstract}

\maketitle


\section{\label{sec:Introduction}Introduction}

The continuous miniaturization of semiconductor devices in pursuit of enhanced performance, energy efficiency, and functional density has created an urgent demand for nanoscale characterization. As device features routinely reach lateral dimensions below 10 nm nowadays \cite{intel2025,Samsung2025,tsmc2025}, traditional electrical analysis techniques \cite{Deen2017} (e.g. Four-Point Probe, Hall measurements) are increasingly reaching their resolution and sensitivity limits. In this context, correlative techniques that combine microscopy with local electrical characterization have become indispensable to understand the behavior of devices and their failure mechanisms on a nanometer scale \cite{Ramesh2024}.

Electron beam-induced current (EBIC) measurements in the scanning electron microscope (SEM) have established themselves as a powerful tool for failure analysis and to map carrier diffusion lengths in semiconductor materials \cite{Leamy1982,Abou-Ras2019}. The focused electron beam creates electron–hole pairs, which are subsequently separated by internal electric fields and collected at the sample contacts, resulting in a measurable current. When integrated with scanning transmission electron microscopy (STEM), EBIC can even reach nanometer resolution \cite{Hubbard2020}, allowing for investigations of internal device structures and interfaces with unprecedented resolution \cite{Conlan2021a,Zutter2021,Saito2021}. High-resolution STEM-EBIC measurements are particularly promising for complex chip architectures such as vertical junctions, heterointerfaces, and dopant gradients that are otherwise challenging to probe.

In this work, we investigate the feasibility and limitations of STEM-EBIC for the quantitative characterization of semiconductor devices. We demonstrate how the sample preparation, using the focused ion beam (FIB), is critical to extracting quantitative information from EBIC measurements. Particular emphasis is placed on understanding how beam-induced preparation artifacts alter the electric properties of a semiconducting TEM sample.

By comparing STEM-EBIC and complementary electrical measurements obtained from TEM lamellae prepared from the same commercial silicon photodiode (PD), with the macroscopic electric properties of the bulk device, we evaluate the potential of the method as a diagnostic tool for next-generation semiconductor devices. Our findings contribute to the growing pool of nanoanalytical methods required to support the semiconductor industry as it advances to sub-5 nm technology nodes, where conventional metrology tools struggle to keep pace \cite{IRDS2024}.

\section{\label{sec:Results}Results and Discussion}

\subsection{\label{sec:Pre-characterisation}Pre-characterisation of the silicon photodiode}

\begin{figure*}[hbt!]
\includegraphics[width=0.9\textwidth]{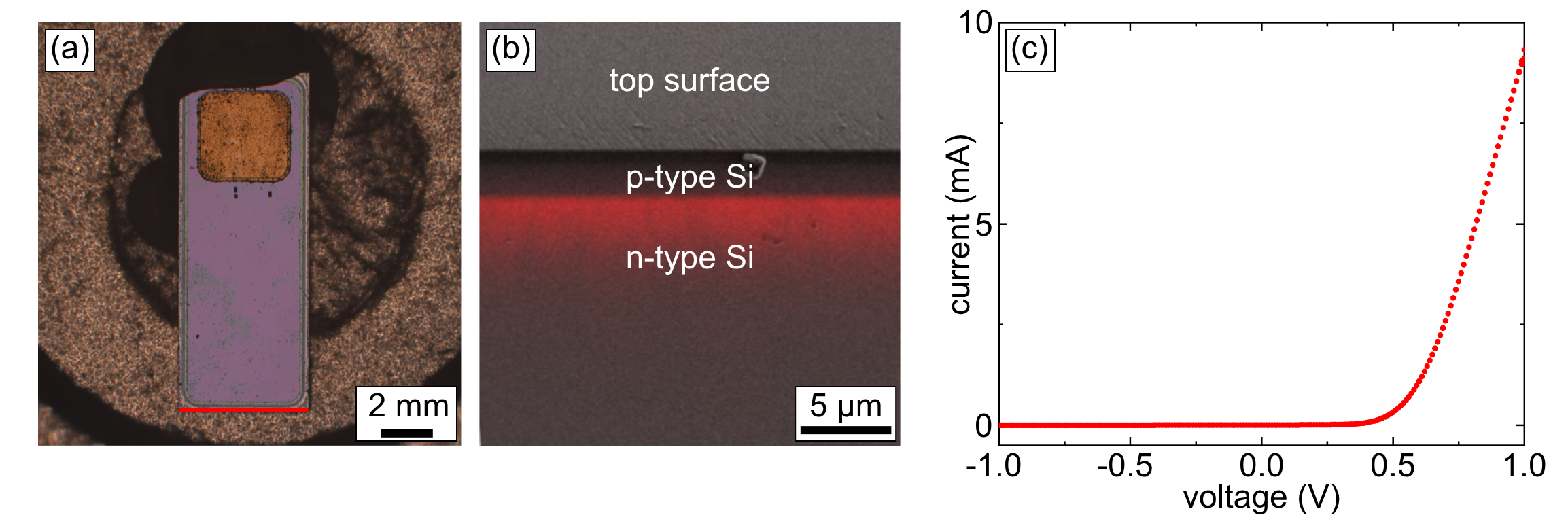}
\caption{\label{fig:Fig. 1}Characteristics of the bulk silicon PD: (a) Light micrograph of the device, with the purple region marking the active area. (b) Secondary electron image of the front side (cf. red line in (a)) overlaid with EBIC contrast (red). (c) Current–voltage characteristics of the PD.}
\end{figure*}

A commercial silicon PD was selected as the sample material because of its simple structure. Additionally, its homogeneous elemental composition makes it particularly suitable for identifying preparation-induced artifacts. A light microscopy image of the PD mounted on an aluminum SEM stub using silver paste is shown in FIG. \ref{fig:Fig. 1} (a). The purple region marks the active area, while electrical contact can be made via the golden solder pad at the top.

To locate the depletion region, the PD was mechanically polished from the front surface (cf. red line in FIG. \ref{fig:Fig. 1} (a))  to reveal its cross-sectional structure. This cross section was subsequently analyzed using SEM-EBIC. FIG. \ref{fig:Fig. 1} (b) shows a secondary electron image overlaid with a red EBIC contrast, revealing the depletion region approximately 1.5 µm beneath the surface.

The electrical characterization was performed by measuring the current–voltage (I-V) characteristics of the PD. For this, platinum probes were used to establish electrical contact with the gold solder pad on the top side and the aluminum stub contacting the backside of the device. The I–V curve, measured with a \textit{Keysight B2902A} source measure unit (SMU), is shown in FIG. \ref{fig:Fig. 1} (c). The measurement shows the typical behavior of a diode with a rectifying response and a forward bias voltage of around 0.6 V, as expected for silicon p–n junctions \cite{Sze2006}. Given the SMU configuration, where 'Force High' corresponds to the gold pad and 'Force Low' to the bottom side of the PD, the measurement confirms that the p-doped layer is situated above the n-doped layer in the PD, as shown in FIG. \ref{fig:Fig. 1} (b).

Since TEM samples possess significantly reduced dimensions, typically requiring thicknesses below 100 nm, which is comparable to the gate pitch in modern devices, their properties can deviate drastically from their bulk counterparts. This ultimately raises the question of whether the previously described characteristics of the standard bulk PD can also be quantified in the TEM samples. To address this issue, the following sections will detail the sample preparation process, electrical characterization, and STEM-EBIC measurements of TEM lamellae prepared from the same silicone PD.

\subsection{\label{sec:Sample preparation}Sample preparation procedures}

Previous studies have shown that sample preparation using a focused ion beam (FIB) can significantly influence the electrical behavior of TEM lamellae in subsequent EBIC measurements. In particular, implantation of gallium ions into surface damage layers can introduce conductive defects, leading to EBIC signals that do not originate from the original device structure \cite{Bonifacio2024a}. To shed more light on this issue, two identical TEM lamellae were prepared and mounted on \textit{Protochips FIB optimized} E-chips. The first was fabricated using gallium ions in an \textit{FEI Helios Nanolab 660} FIB (Ga-FIB), and the second using xenon ions in an \textit{FEI Helios G4} Plasma FIB (PFIB), avoiding metallic contamination \cite{Hubbard2024}.

The samples were prepared as follows: (i) deposition of a protective platinum layer on the PD surface using the Ga-FIB; (ii) extraction of two 2.5 µm-thick blanks for lamella fabrication with the Ga-FIB; (iii) attachment to commercial copper TEM grids via Ga-FIB welding; (iv) thinning to 800 nm using the Ga-FIB and PFIB, respectively; (v) cutting, transfer, and welding of the lamellae onto the E-chip with electron-beam-deposited tungsten; (vi) final thinning to 500 nm with the Ga-FIB and PFIB, respectively, and (vii) slitting of the p-n junction to selectively connect the n-type region to the left and the p-type region to the right contact pad of the E-chip. 

Since both samples exhibit amorphous damage layers after the FIB preparation, additional low-energy (900 eV) argon ion milling was carried out in a \textit{Fischione Instruments NanoMill} to remove these layers and further thin the specimens \cite{Bonifacio2024b}. Given the dependence of the EBIC signal strength on the sample thickness \cite{Conlan2021a} and the ongoing debate regarding a minimum thickness for signal detection, EBIC measurements of both samples after each milling step were systematically complemented by a quantitative thickness determination and compositional analysis using STEM-EELS and -EDX. 

\begin{figure}[hbt!]
\includegraphics[width=0.5\textwidth]{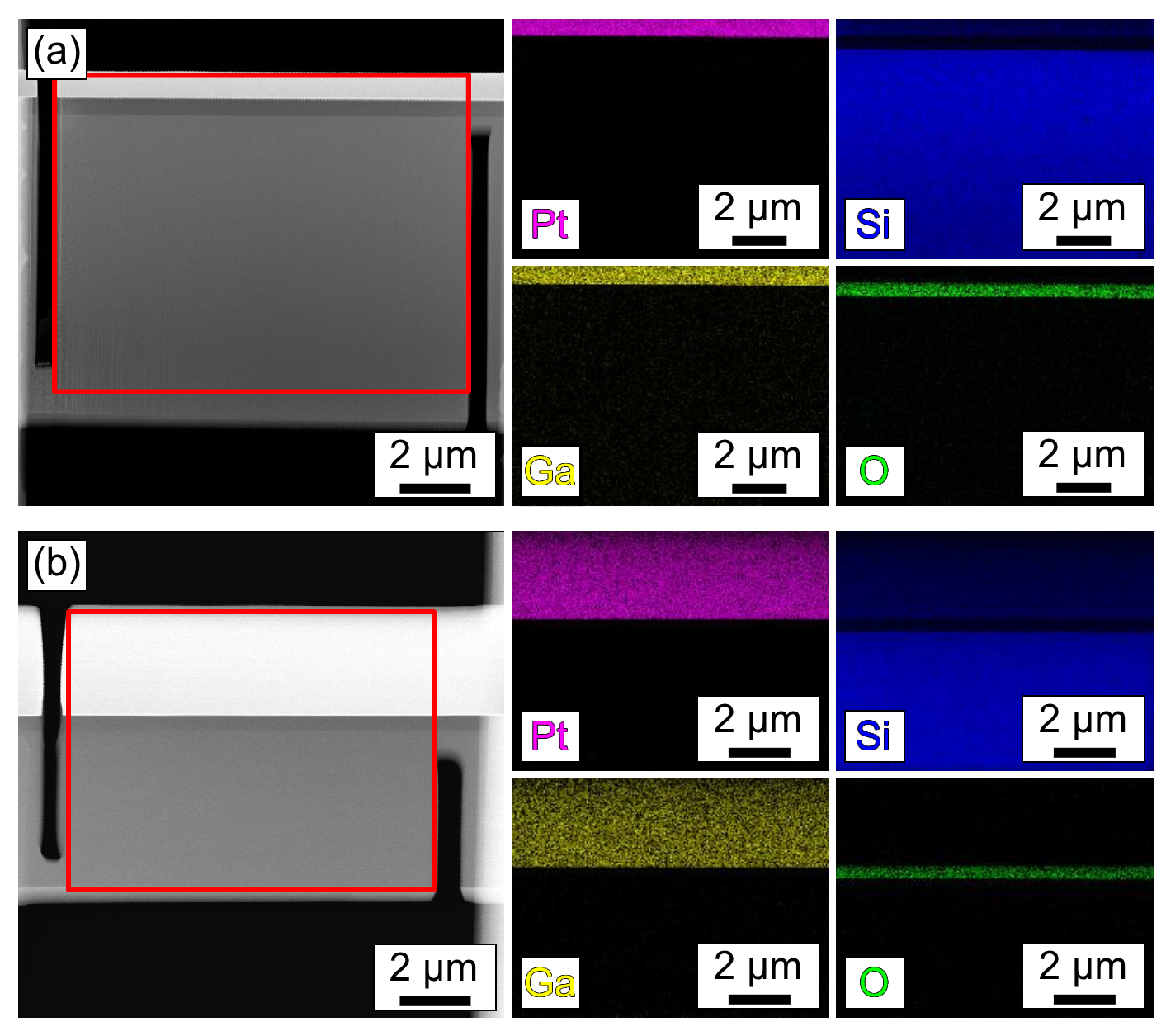}
\caption{\label{fig:Fig. 2}STEM measurements. Left column: ADF images of the two TEM samples prepared using Ga-FIB (a) and PFIB (b). Right columns: Corresponding elemental maps of the regions highlighted by red rectangles in the left column.}
\end{figure}

STEM-EDX measurements of the two samples are shown in FIG. \ref{fig:Fig. 2}. From top to bottom, the cross section reveals the protective platinum–carbon layer, which in both cases contains gallium due to the use of the Ga-FIB during blank preparation. Underneath lies a passivating silicon oxide layer, followed by the silicon region that hosts the p-n junction of the PD (cf. FIG. \ref{fig:Fig. 1} (b)).

\subsection{\label{sec:I-V characteristics}Current–voltage characteristics}

\begin{figure}[hbt!]
\includegraphics[width=0.45\textwidth]{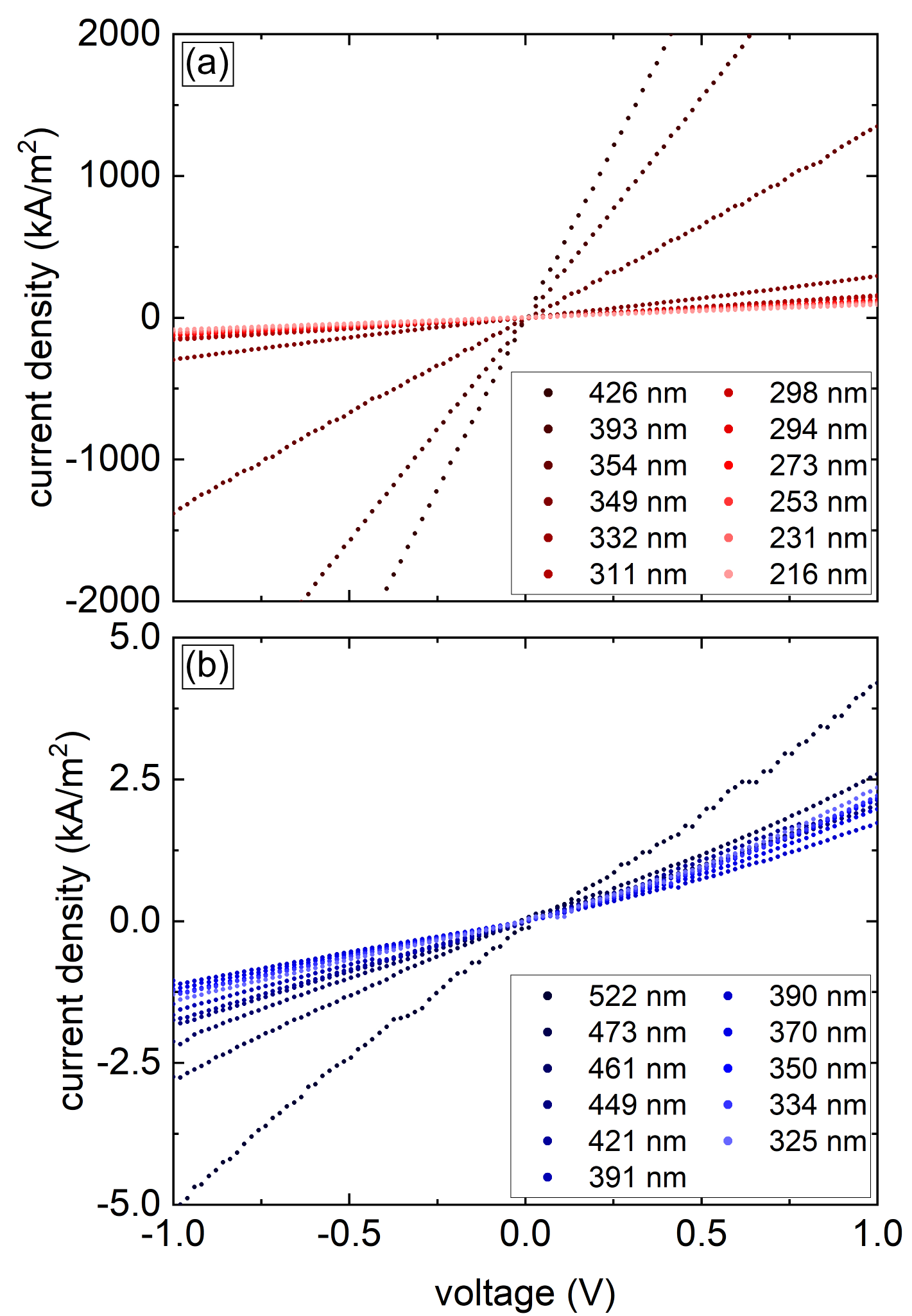}
\caption{\label{fig:Fig. 5}I–V characteristics of the TEM lamellae prepared using the Ga-FIB (a) and PFIB (b). The mean sample thicknesses, averaged over the whole silicon area between the slits, are indicated in the legend. Note that the current density range is larger by three orders of magnitude in (a) with respect to (b).}
\end{figure}

The j–V characteristics of both TEM lamellae were recorded using a \textit{Point Electronic Electrical Analysis Amplifier}. For each thickness step, the currents were measured in the voltage range $-1 \, \mathrm{V} \le U \le 1 \, \mathrm{V}$, as shown in FIG. \ref{fig:Fig. 5}.

For the Ga-FIB sample, the initial j–V curves exhibit a low-resistance, nearly ohmic behavior, which can be attributed to the gallium implantation producing highly conductive metallic surface layers (cf. FIG. \ref{fig:Fig. 5} (a)). During subsequent argon ion milling, the ion beam was scanned across the lamella between the slits to preserve the electrical contact, progressively removing the amorphous layers and resulting in an increase in resistance. After approximately a total of 70 nm of material was removed from the lamella center, the j–V characteristics stabilized, with no significant changes observed for thicknesses below 349 nm. Nevertheless, even after removal of these implanted layers in the central region, the typical diode-like I–V characteristics could not be recovered.

Examining the j–V characteristics of the PFIB lamella, one notices that its conductivity is about three orders of magnitude lower than that of the Ga-FIB sample. This difference can be attributed to the absence of gallium implantation. After thinning the sample by roughly 50 nm, no substantial changes are observed in the measurement curves anymore. However, the curves also do not exhibit the typical shape of a diode characteristic (cf. FIG. \ref{fig:Fig. 5} (b)).

Interestingly, the course of the PFIB j-V curve is S-shaped rather than linear. Such a behavior is characteristic of Schottky barriers \cite{Conlan2021b} and suggests why neither sample shows the characteristics of a diode. It is likely that high-resistance Schottky barriers form at the silicon–platinum welding interfaces, causing a significant portion of the applied voltage to drop across these contacts rather than across the p-n junction. Consequently, the effective voltage reaching the p–n junction remains well below the threshold voltage, preventing the sample to become highly conductive typical for diodes.

\subsection{\label{sec:STEM EBIC}STEM-EBIC measurements}

Although the electrical behavior of the TEM lamellae differs from the expected diode characteristic, STEM-EBIC measurements were carried out on both samples to assess whether they could be quantitatively analyzed using this technique. STEM-EBIC measurements were performed at each sample thickness for both lamellae using a \textit{point electronic Electrical Analysis Amplifier}. The left contacts in FIG. \ref{fig:Fig. 2}, connecting the n-doped region to the E-chip, was grounded, while the current generated in the p-type silicon was routed through the right contact into the amplifier. 

\begin{figure}[hbt!]
\includegraphics[width=0.5\textwidth]{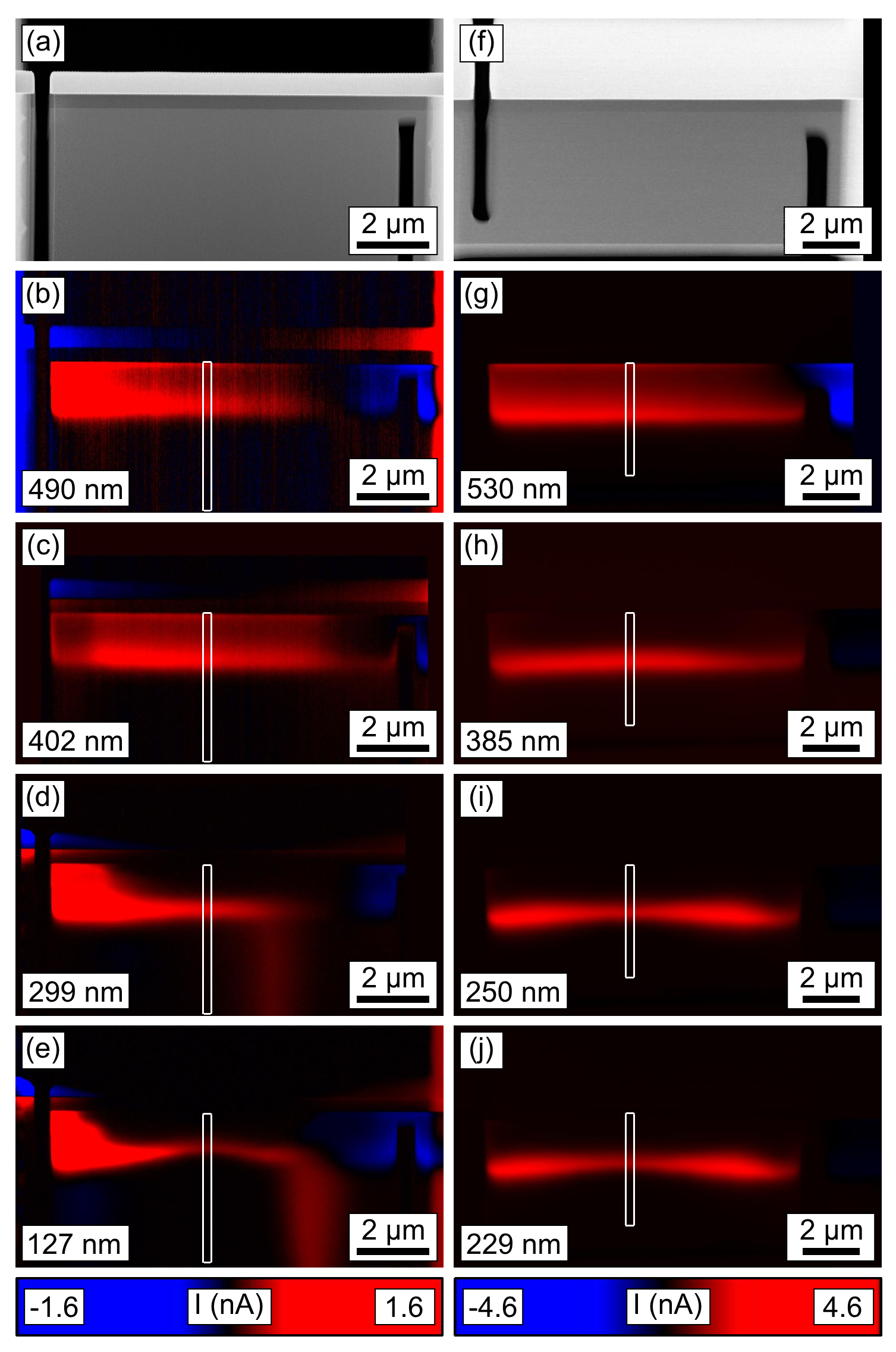}
\caption{\label{fig:Fig. 3}STEM-EBIC maps of the Ga-FIB (b) - (e) and PFIB (g) - (j) samples, with the depletion region visible as red contrast. The white rectangles mark the positions of the linescans shown in Fig. \ref{fig:Fig. 4}. Sample thicknesses were determined at the points where the linescans intersect the depletion region. The ADF images of the corresponding sample areas for the Ga-FIB and PFIB are displayed in (a) and (f).}
\end{figure}

Representative quantitative EBIC images are shown in FIG. \ref{fig:Fig. 3}. Since electron-hole pairs that are created in the depletion region around the p-n junction are separated by the local electric field of the space charges and thus contribute to a net current across the junction in both the Ga-FIB and PFIB samples, the depletion region appears as a red contrast about 1.2 µm below the oxide layer, consistent with the SEM-EBIC mesaurement in Fig. \ref{fig:Fig. 1}(b). In the PFIB lamella (cf. FIG. \ref{fig:Fig. 3} (b)), the EBIC map shows a uniform red contrast between the slits, while the Ga-FIB sample (cf. FIG. \ref{fig:Fig. 3} (a)) reveals a gradient that transitions from red to blue across the lamella from left to right.

Particular attention should be paid to the evolution of the EBIC signal during the milling process. In the as-prepared Ga-FIB and PFIB lamellae, the EBIC contrast extends all the way from the depletion region up to the oxide layer. However, as the samples are thinned, the signal persists only in the local environment of the p-n junction and becomes more sharply defined. Because this effect is observed in both samples, it can only be attributed to the amorphous surface layer produced during FIB milling, which forms a heterojunction with the underlying crystalline silicon. At such an amorphous/crystalline silicon interface, internal electric field arise from the difference between the band gaps of the two materials \cite{Fuhs1974}, an effect that is commercially exploited in heterojunction solar cells \cite{Pankove1979,Taguchi2005}. The EBIC signal is more pronounced on the (bottom) p-type side because the minority carriers, electrons, exhibit a higher mobility and longer diffusion lengths than holes in n-type Si \cite{Conlan2021a}, enabling a more efficient collection. Once the amorphous layer is removed, this signal vanishes.

\begin{figure*}[hbt!]
\includegraphics[width=0.9\textwidth]{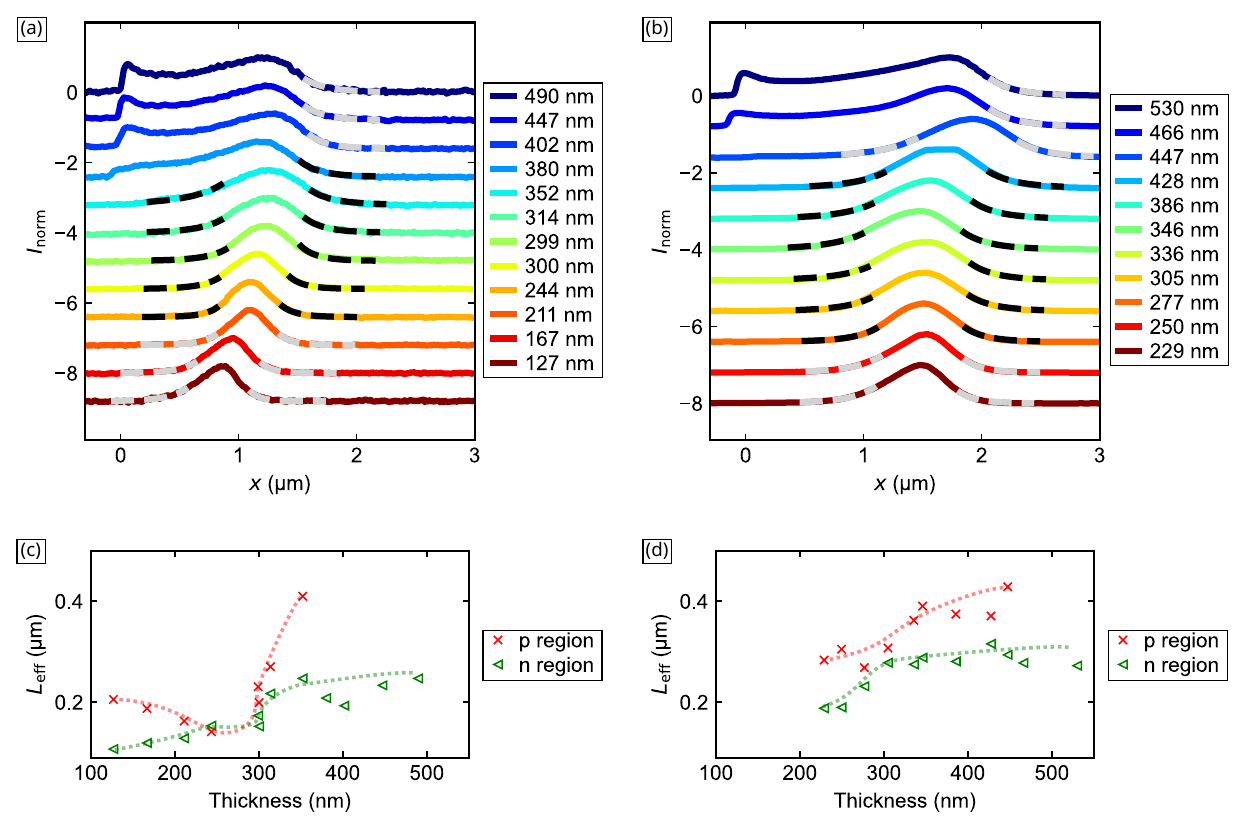}
\caption{\label{fig:Fig. 4}Line profiles of the EBIC signal and effective diffusion lengths: the line profiles of the normalized EBIC signal for the indicated sample thicknesses measured on samples prepared with Ga-FIB (a) and PFIB (b).The curves were vertically offseted for better visibility. The gray and black dashed lines indicate fits of Eq. \ref{eq:diffusion length} to the rising and falling edges of the peaks (see text for details), i.e. on the p- and n-type side of the junction, respectively. (c) and (d) Effective diffusion for electrons (red) and holes (green) as obtained from fits of Eq. \ref{eq:diffusion length} in the p- and n-regions for the Ga-FIB (c) and PFIB (d) samples, respectively.}
\end{figure*}

To gain further insight, line profiles were extracted from the EBIC maps (cf. white rectangles in Fig. \ref{fig:Fig. 3}). The profiles were taken at the lamella center, where material removal during the argon ion milling is most pronounced. The resulting profiles are shown in FIGs. \ref{fig:Fig. 4} (a) and (b). At the initial thicknesses, both samples exhibit a double-peak structure. Upon thinning, the first peak next to the oxide layer decreases successively and disappears after removal of approximately 70 nm and 40 nm thick surface layers in the Ga-FIB and PFIB samples, respectively, which is consistent with previous publications on FIB-induced surface amorphization \cite{Kelley2013}.    

These profiles allow to estimate the effective diffusion lengths of the holes and electrons created by the electron beam in the PD. The EBIC profile of a p-n junction decays approximately as \cite{Ong1994}:

\begin{equation}{\label{eq:diffusion length}}
I(x) = k \cdot x^{\alpha} \cdot e^{-\frac{x}{L_{eff}}},
\end{equation}

where $x$ denotes the distance from the center of the depletion region and $L_{eff}$ the effective diffusion length of the minority carriers. The exponent $\alpha$ accounts for surface recombination, taking values of 0 and $-\frac{1}{2}$ in the case of negligible and strong surface recombination, respectively \cite{Duchamp2020}, and $k$ is a scaling parameter.

For the analysis of the profiles, the baseline, as determined approximately 1 µm below the depletion region, was first subtracted from the experimental data. The profiles were then normalized to their maxima, and the peak positions together with the two deflection points at the rising and falling edges of the peak were determined. The lower and upper flanks of the peaks were subsequently fitted by Eq. \ref{eq:diffusion length} within a range of $\pm$ 0.05 µm to $\pm$ 0.7 µm from the deflection point on either side. In these fits, the origin was fixed at the peak position, $k$ was set to 1, and $\alpha$ to $-\frac{1}{2}$, reflecting the surface modification induced by the FIB milling. The resulting fits are shown as dashed lines superposed the experimental data in the top row of Fig. \ref{fig:Fig. 4}. The hereby extracted effective diffusion lengths are summarized as a function of sample thickness in FIG. \ref{fig:Fig. 4}(c) and (d), which are significantly shorter than previously reported bulk values \cite{Ioannou1982,Marcelot2019}. While the bulk diffusion length is governed by the carrier lifetime and diffusion coefficient, charge carrier recombination in unpassivated and/or damaged surface layers of thin samples provides an additional loss channel, thereby reducing the effective diffusion length\cite{Lahreche2022}.

\section{\label{sec:Conclusion}Conclusion}

We systematically investigated the electrical properties of two TEM lamellae prepared from a silicon photodiode using Ga-FIB and PFIB, and compared them to those of the bulk device. Using STEM-EBIC, the p–n junctions in the thin samples could be clearly identified. Preparation artefacts, most pronounced in the Ga-FIB lamella, caused a blurring of the EBIC signal, but were effectively removed by subsequent argon ion milling. Quantitative analysis of the EBIC profiles across the p-n junctions allowed us to extract the effective carrier diffusion lengths of both holes and electrons as a function of local sample thickness. The measured diffusion lengths are up to three orders of magnitude smaller than those obtained from SEM-EBIC measurements on bulk silicon, underscoring the dominant influence of surface recombination and FIB-induced surface modifications on the EBIC contrast in thin lamellae.

In addition to the diffusion length analysis, the j–V characteristics of the TEM samples deviate strongly from those of the bulk PD. Even after the argon ion milling, which removes conductive surface layers, neither the Ga-FIB nor the PFIB lamella exhibited typical diode behavior. We attribute this to high-resistance contacts formed during the welding of the semiconductor to the metallic carrier, suggesting that proper work-function matching between the device and the contacts is essential \cite{Yoshitake2021}. Preliminary compositional analysis of the gold contact pad already provides insight into the layer structure, but a robust solution would require the fabrication of customized chips with conductor tracks adapted to the respective sample.

Overall, the presented methodology demonstrates the potential of STEM-EBIC for quantitative studies of more complex device architectures. Beyond fundamental investigations of carrier transport, it offers a pathway to optimize nanoscale device performance by directly correlating microstructural features such as dislocations or interfaces with their electronic activity.

\begin{acknowledgments}
This research was supported by the Deutsche Forschungsgemeinschaft (DFG) through TRR 404 Active-3D (project number 528378584). We also gratefully acknowledge Bernd Rellinghaus for his valuable discussions during the preparation of the manuscript.
\end{acknowledgments}

\appendix

\nocite{*}

\bibliography{references}

\end{document}